\documentclass{mem}
\usepackage{natbib}\usepackage{txfonts}\usepackage{balance}
\usepackage{graphicx}
\usepackage[a4paper,breaklinks,dvipdfm]{hyperref}
\idline{0}{0}
\begin{document}
\def\teff{$T\rm_{eff }$}
\def\kms{$\mathrm {km s}^{-1}$}

\title{
Finding Hot-Jupiters by \textit{Gaia} photometry using the Directed Follow-Up strategy
}

   \subtitle{}

\author{
Y. \,Dzigan\inst{1} 
\and S. \,Zucker\inst{2}
          }

  \offprints{Y. Dzigan}

\institute{
Particle Physics and Astrophysics Dept.,
Weizmann Institute of Science, 
Rehovot, 76100, Israel
\email{yifatdzigan@gmail.com}
\and
Geophysics, atmospheric and Planetary Sciences Dept., 
Tel Aviv University, Tel Aviv 69978, Israel
}

\authorrunning{Dzigan}

\titlerunning{Directed Follow-Up strategy for \textit{Gaia} photometry}

\abstract{
All-sky surveys of low-cadence nature, such as the promising {\it Gaia} Space mission, have the potential to "hide" planetary transit signals. We developed a novel detection technique, the Directed Follow-Up strategy (DFU), to search for transiting planets using sparse, low-cadence data. 
According to our analysis, the expected yield of transiting Hot-Jupiters that
can be revealed by Gaia will reach a few thousands, if the DFU strategy will be applied to facilitate detection of transiting planets with ground-based observations.
This will guaranty that \textit{Gaia} will exploit its photometric capabilities and will have a strong impact on the field of transiting planets, and in particular on detection of Hot-Jupiters. 
Besides transiting exoplanets \textit{Gaia}'s yield is expected to include a few tens of transiting brown dwarfs, that will be candidates for detailed characterization, thus will help to bridge the gap between giant planets and stars.
}
\maketitle{}

\section{Introduction}

Transiting planets, in particular Hot-Jupiters (HJs) orbiting bright stars, are of the most appealing objects to detect, since they are favorable for observational follow-up study.  As such, they are keys for studying the mechanism that drive planetary formation, migration and evolution. 
The Directed Follow-Up strategy (DFU) is a detection approach that we use to search for transiting HJs in sparse, low-cadence data, such as \textit{Hipparcos} \citep{ESACat1997}, and its successor {\it Gaia} \citep{2012Ap&SS.tmp...68D}.
Several studies examined the feasibility of
detecting transits in the \textit{Hipparcos} Epoch
Photometry, and made \textit{posterior} detections of two known transiting planets in \textit{Hipparcos}, using 
the previously available knowledge of the orbital elements of the
exoplanets \citep[e.g.,][]{Robichon2000, 2005A&A...444L..15B}. Those studies concluded that \textit{Hipparcos}
was not sufficient for transit detection,
without using prior information \citep{2006A&A...445..341H}. The posterior detections prove that the information about the transits is buried in the data, and that motivated us to look for a way to utilize low-cadence photometric surveys for transit search.

\section{Directed Follow-Up strategy for low-cadence surveys} 

The DFU strategy is based on Bayesian inference that we use to estimate the posterior probability density functions of the
transit parameters. We describe the transit by a simple box-shaped
light-curve \citep[.e.g,][]{2002A&A...391..369K} with five parameters: the period -- $P$, phase -- $T_c$, and width of the
transit -- $w$, and the flux levels in and out of transit. 

First we apply a Metropolis-Hastings (MH)
algorithm (implementation of a Markov-Chain Monte-Carlo procedure) to the measurements of a target
star. The results (after excluding the appropriate `burning time') are
Markov chains of the successful iterations, for each of the model
parameters. From each chain we extract the stationary distribution, which we use as the parameter's estimated Bayesian posterior
distribution \citep{2005blda.book.....G}.  If the low-cadence data
happen to sample enough separate transits, with sufficient
precision, we expect the distributions to concentrate around the
solution of the transit. But unlike the case of precise high-cadence surveys, even if the star hosts a transiting planet, due to the low-cadence observations, only a few transits will be sampled. Therefore, the distributions will probably spread over
different solutions, besides the unknown actual one.

The second step is to assign probability for a transit for future times, which is represented by the Instantaneous Transit Probability function (ITP).
Then we will choose stars for a follow-up campaign according to the ITP peak values and skewness, both are indications for the presence (or absence) of a periodic transit-like signal in the data. Another criterion is the Wald statistic of the transit
depth posterior distribution, that quantifies the significance
of the transit depth by measuring it in terms of its own standard
deviation.  

The last step is to actually perform
follow-up observations, at the times directed by the ITP, thus
optimizing the chances to detect the transit in a few observations as
possible. Then the follow-up observations should be combined with the  data from the
survey, to recalculate new posterior
distributions that reflect our new state of knowledge, and to propose
new times for the next follow-up observations.

If a follow-up observation happens to take place during transit, it will usually eliminate most
spurious peaks in the period posterior distribution (PPD), except for
the actual orbital period of the planet. 
However, usually we will \textit{not} observe the
transit in the follow-up observations. In this case the new data will still eliminate some periods that will not fit our new state of knowledge, resulting in modified posterior distributions of the model
parameters.  The whole procedure should be repeated until the
detection of a transiting planet, or, alternatively the exclusion of
its existence.
After we will detect a planetary candidate a careful high-cadence photometric or spectroscopic follow-up can confirm its planetary nature.  

\section{DFU application to \textit{Gaia}}

\textit{Gaia} was initially intended to detect exoplanets using its unprecedented astrometry, and in this aspect it will 
have an innovative contribution to planetary research \citep[see e.g.,][]{2010ASPC..430..253L, 2013EPJWC..4715005S, Sozzetti14}. 
In \cite{2012ApJ...753L...1D} we showed that despite the non-competitive photometry of the telescope (relative to dedicated transit surveys), \textit{Gaia} will be useful for transit search. 
In \cite{2013MNRAS.428.3641D} we demonstrated the application of the strategy to \textit{Gaia} using a few simulated cases,and advocate for the use of real-time follow-up resources as means to identify transiting planets, to begin during the operational lifetime of the mission.

\subsection{\textit{Gaia}'s expected yield of transiting exoplanets}

\begin{figure}[]
\resizebox{\hsize}{!}{\includegraphics[clip=true]{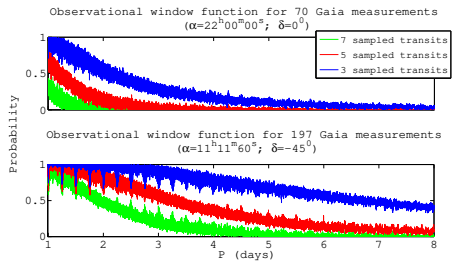}}
\caption{
\footnotesize
Observational window functions for two areas of
\textit{Gaia}'s sky, as discussed in \citet{2013MNRAS.428.3641D}.
\textit{Top}: A sky direction that \textit{Gaia} is expected to visit
$70$ times. \textit{Bottom}: A sky
direction that \textit{Gaia} is expected to sample $197$ times. Both
window functions were calculated for a transit duration of
$2\,\mathrm{hr}$, for a minimum of three, five and seven in-transit
observations.}
\label{fig.window_funcion1}
\end{figure}

To estimate \textit{Gaia}'s expected yield of transiting exoplanets, we followed a statistical methodology \citep{2008ApJ...686.1302B} that accounts for the transit probability, transiting planet frequencies based on complete transit surveys (namely OGLE), and assumptions regarding the galactic structure. We found that the potential yield can reach a few thousands, depending on the
number of planetary transits that the telescope should sample to
secure detection, and therefore on the detection algorithm \citep{2012ApJ...753L...1D}. The observational window functions \citep{2009ApJ...702..779V} in Fig.~\ref{fig.window_funcion1} represent the probability to sample a minimum of three, five and seven transits for a typical
\textit{Gaia} star, with $70$ measurements (upper panel), together with another, less
probable case, with $197$ measurements (bottom panel). 
For a typical \textit{Gaia} star that hosts a transiting HJ (upper panel of Fig.~\ref{fig.window_funcion1}), the probability to sample at least seven separate transits is practically negligible. The probability to sample five transits is $\sim5\%$, while the
probability to sample a minimum of three transits increases to $\sim30\%$. 
Thus, we conclude that it will be beneficial to relax
the requirement for a minimum number of sampled transits that can secure a detection, and the DFU
strategy aims to achieve exactly that. It is worth nothing that for stars located at ecliptic latitudes of $\pm45$ deg, the probability to sample at least three different transits amount to almost $100\%$ (bottom panel of Fig.~\ref{fig.window_funcion1}).

\subsection{application to Gaia}

We examined the application of the strategy to \textit{Gaia} using a few simulated cases, inspired by known transiting planets. 
We simulated light curves of planetary transits according to
\textit{Gaia}'s scanning-law and expected photometric precision \citep{2013MNRAS.428.3641D}. 
Each planet was assigned with a transit epoch that implied different number of transit samples.
We found that DFU can result in detection
through one of two main scenarios.
The \textit{first scenario}
is a detection using \textit{Gaia}
data alone. In this scenario the MH algorithm
culminates in a PPD that is centered around a single solution, (and maybe
its harmonics and sub-harmonics). The compact distribution will
guarantee that the resulting ITP peaks will coincide with future 
transits. Therefore we will be able to direct a high cadence follow-up
observations, to examine the nature of the candidate.
We found that in case \textit{Gaia} will sample five different transits pwe system, the strategy will result in a detection using \textit{Gaia} data alone (an example can be found in \cite{2013MNRAS.428.3641D}).

Due to \textit{Gaia}'s low cadence, in most cases we expect to obtain
multimodal PPD \citep{2011MNRAS.415.2513D}. This will result in the \textit{second scenario}, in which we will require more
than a single follow-up observation in order to finally detect the
transit. 

\begin{figure}[]
\resizebox{\hsize}{!}{\includegraphics[clip=true]{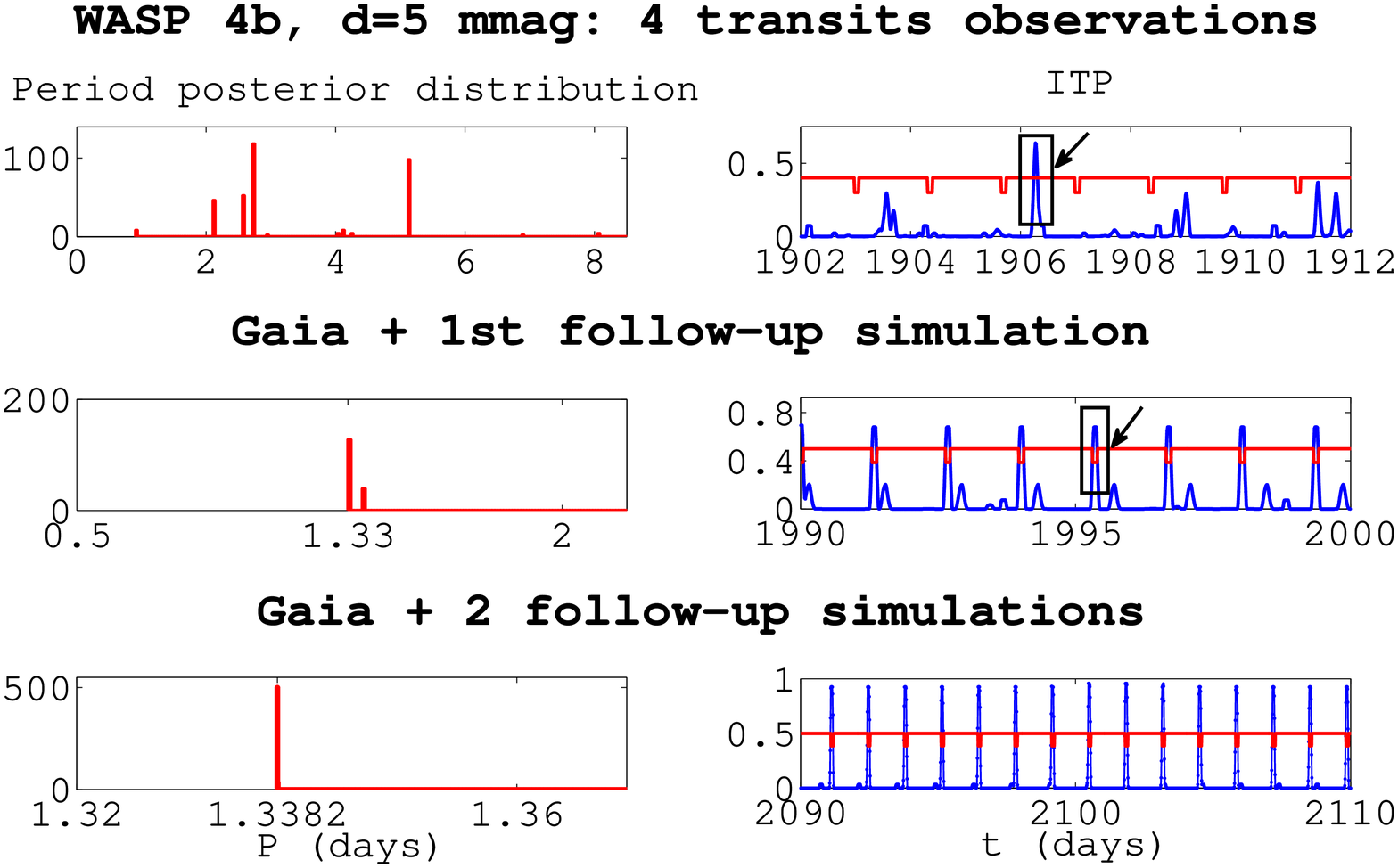}}
\caption{
\footnotesize
Application of the DFU strategy for \textit{Gaia} simulation, as discussed in \citet{2013MNRAS.428.3641D}.
\textit{Top}: PPD and ITP for a simulation inspired by WASP-4b, $d=5\,\mathrm{mmag}$, 4 sampled transits. 
\textit{Middle}: PPD and ITP for the combined datasets (adding the first follow-up observing sequence, simulated according to the marked ITP peak, {\bf not} during transit.)
\textit{Bottom}: New PPD and ITP after adding the second follow-up sequence, which was simulated according to the marked ITP peak that this time did sample the hypothetical transit.}
\label{fig.wasp4_fu}
\end{figure}

We demonstrate the strategy application for \textit{Gaia} in Fig.~\ref{fig.wasp4_fu}, using a simulation inspired by WASP-4b. We tailored the transit phase, so that \textit{Gaia}'s scanning-law will sample four individual transits during the complete mission lifetime (five nominal years). For a relatively small transit depth (five mmag) the strategy
yield a detection using two directed follow-up observations. We found that with as little as three transits samples, we would be able to trigger follow-up observations that will eventually lead to the transit detection (using two or more follow-up sequences).

\subsection{Alerts for follow-up observations during \textit{Gaia} operation}
Due to the ITP degradation \citep{2011MNRAS.415.2513D}, we suggest that the follow-up campaign will begin as soon as possible, while \textit{Gaia} still operates. 
In \cite{2013MNRAS.428.3641D} we examined simulations using half of \textit{Gaia}'s timespan, and found it will be possible 
to trigger follow-up observations if \textit{Gaia} will sample three transits in this short time interval (at least two years of data).  
  
The DFU strategy requires a widely spread network, able to
follow-up on prominent ITP peaks during the mission lifetime. 
\textit{Gaia} Science Alerts team \citep{2012IAUS..285..425W} is assigned to trigger follow-up observations for
transient events in the \textit{Gaia} data stream.  Although planetary transits are not transient, the information about the transit is, so Science Alerts may prove as beneficial for the strategy implementation.

\section{Discussion and Summary}

We introduced the DFU strategy, and examined its application to \textit{Gaia}
photometry. 
We found that for all simulated light curves 
with transits deeper than $1\,\mathrm{mmag}$, the DFU strategy can be used to either 
recover the periodicity, or to propose times for directed 
follow-up observations that will lead to detection. 
If a typical HJ ($d \sim 10\,\mathrm{mmag}$), will be sampled by \textit{Gaia} 
during five different transits, a detection will be possible using \textit{Gaia} data alone. 
A more realistic scenario, with only three in-transit measurements, will usually result 
in a detection under a follow-up campaign, based on the ITP most significant peaks, 
that will reduce the observational efforts to a minimum. Test light curves with no transit signal were never classified as
candidates, and were not prioritized for follow-up observations. 

We showed \citep{2012ApJ...753L...1D} that the false-alarm rate due
to \textit{Gaia}'s white Gaussian noise is negligible, and that our prioritization procedure eliminated all the false alarms that resulted from stellar red noise \citep{2013MNRAS.428.3641D}. 

In \cite{2012ApJ...753L...1D} we studied the yield of transiting planets from \textit{Gaia}, and predict that the yield can reach a few thousands, provided that the detection algorithm will be tailored for \textit{Gaia}'s special features.  

Since our ability to schedule follow-up observations based on the ITP degrades with time, the optimal scenario for \textit{Gaia} will be to initiate the follow-up campaign while \textit{Gaia} still operates. Simulating only two to three years of \textit{Gaia} observations, we found that we will be able to prioritize stars and trigger follow-up observations using partial data \citep{2013MNRAS.428.3641D}. As indicated by the window function in Fig.~\ref{fig.window_funcion1}, stars located close to the knots of the scanning-law will be sampled during different transits more quickly, thus allowing the follow-up to begin even earlier.

In addition to transiting HJs, \textit{Gaia} is expected to yield a few tens of transiting brown dwarfs \citep{Bouchy14}, that can be studied in details using follow-up observations, and will help to bridge the gap between giant planets and stars.

We conclude that
\textit{Gaia} photometry, although not optimized for transit detection, should
not be ignored in the search of transiting planets. 
A timely application of the DFU strategy will eventually yield the detection of thousands of transiting HJs, and will promise that \textit{Gaia} will contribute to the photometric search of planets via transits, along with its anticipated astrometric detections.
 
\begin{acknowledgements}
We are grateful to Laurent Eyer and all DPAC-CU7 team for their insights regarding \textit{Gaia} photometry. 
\end{acknowledgements}

\bibliographystyle{aa}

\begin{thebibliography}{}

\bibitem[Beatty \& Gaudi(2008)]{2008ApJ...686.1302B}
Beatty, T.~G.~\& Gaudi, B.~S.\ 2008, \apj, 686, 1302

\bibitem[Bouchy et al.(2005)]{2005A&A...444L..15B}
Bouchy, F. et al.\
2005, \aap, 444, L15

\bibitem[Bouchy(2014)]{Bouchy14} Bouchy, F.\ 2014, \memsai, xx

\bibitem[de Bruijne(2012)]{2012Ap&SS.tmp...68D}
de Bruijne J.~H.~J.\ 2012, \apss, 68

\bibitem[Dzigan \& Zucker(2011)]{2011MNRAS.415.2513D}
Dzigan, Y.~\& Zucker, S.\ 2011, \mnras, 415, 2513

\bibitem[Dzigan \& Zucker(2012)]{2012ApJ...753L...1D} 
Dzigan, Y.~\& Zucker, S.\ 2012, \apjl, 753, L1 

\bibitem[Dzigan \& Zucker(2013)]{2013MNRAS.428.3641D} 
Dzigan Y.~\& Zucker, S.\ 2013, \mnras, 428, 3641 

\bibitem[ESA(1997)]{ESACat1997}
ESA, 1997, The Hipparcos and Tycho Catalogues, ESA SP-1200


\bibitem[Gregory(2005)]{2005blda.book.....G}
Gregory, P.~C.\ 2005, Bayesian Logical Data Analysis for the Physical Science, Cambridge University Press

\bibitem[H{\'e}brard \& Lecavelier Des Etangs(2006)]{2006A&A...445..341H}
H{\'e}brard, G.~\& Lecavelier Des Etangs, A.\ 2006, \aap, 445, 341


\bibitem[Kov{\'a}cs et al.(2002)]{2002A&A...391..369K}
Kov{\'a}cs, G.~et al.\ 2002, \aap, 391, 369

\bibitem[Lattanzi \& Sozzetti(2010)]{2010ASPC..430..253L} 
Lattanzi, M.~G.~\& Sozzetti, A.\ 2010, Pathways Towards Habitable Planets, 430, 253 

\bibitem[Robichon \& Arenou(2000)]{Robichon2000} 
Robichon, N.~\& Arenou, F.\ 2000, \aap, 355, 295 

\bibitem[Sozzetti(2013)]{2013EPJWC..4715005S} Sozzetti, A.\ 2013, EPJ Web of Conferences, 47, 15005 

\bibitem[Sozzetti(2014)]{Sozzetti14} Sozzetti, A.\ 2014, arXiv:1406.1388

\bibitem[von Braun et al.(2009)]{2009ApJ...702..779V}
von Braun, K.~et al.\ 2009, \apj, 702, 779
	
\bibitem[Wyrzykowski \& Hodgkin(2012)]{2012IAUS..285..425W}
Wyrzykowski, L.~\& Hodgkin, S.\ 2012, IAU Symp. 285, 425


\end{thebibliography}

\end{document}